\documentstyle[seceq,graphicx,amstext]{jpsj}

\twocolumn

\newcommand{\tJ}{$t$-$J$\ }

\newcommand{\cP}{{\mathcal P}}

\newcommand{\BEQ}[1]{\begin{equation}\label{#1}}
\newcommand{\eeq}{\end{equation}}
\newcommand{\EQA}[1]{\begin{eqnarray}\label{#1}}
\newcommand{\PRL}[1]{Phys. Rev. Lett. {\bf #1}}
\newcommand{\PRB}[1]{Phys. Rev. B {\bf #1}}

\DeclareGraphicsExtensions{.eps} 
\graphicspath{{figures/}{./}}
\newcommand{\myfig}[1]{\begin{figure}[tb]
			  \begin{center} 
				\includegraphics*[width=\linewidth]{#1}
                        \end{center}}

\def\runtitle{Effect of the Orbital Level Difference in Doped Spin-$1$ Chains}
\def\runauthor{Beat {\sc Ammon} and Masatoshi {\sc Imada}}

\title
{
Effect of the Orbital Level Difference in Doped Spin-$1$ Chains
}

\author
{ 
Beat {\sc Ammon} \footnote{E-mail address: ammon@ginnan.issp.u-tokyo.ac.jp}
and Masatoshi {\sc Imada} \footnote{E-mail address: imada@issp.u-tokyo.ac.jp}
}

\inst
{
ISSP, University of Tokyo, 7-22-1 Roppongi, Minato-ku, Tokyo 106-8666, Japan
}

\recdate
{
\today
}

\abst
{ Doping of a two-orbital chain with mobile $S=1/2$ Fermions and
strong Hund's rule couplings stabilizing the $S=1$ spins strongly
depends on the presence of a level difference among these orbitals. By
DMRG methods we find a finite spin gap upon doping and dominant
pairing correlations without level-difference, whereas the presence of
a level difference leads to dominant charge density wave (CDW)
correlations with gapless spin-excitations. The string correlation
function also shows qualitative differences between the two models. 
}

\kword
{
spin-$1$ systems, orbital level-difference, Mott-Hubbard transition,
superconductivity, string-order
}
\begin{document}
\maketitle

Hole doping of strongly coupled spin-liquid groundstates is of great
importance for the understanding of the Mott-Hubbard transition. A
very special spin-liquid is the antiferromagnetic (AF) $S=1$
Heisenberg (HB) chain, which shows a finite spin gap \cite{Haldane}
and a hidden topological order, the string-order parameter
\cite{StringOrder}. Only recently it has become possible to study 
experimentally the effect of doping in a spin-1 chain with mobile
holes in the system $\rm Y_{2-x}Ca_xBaNiO_5$ \cite{YBaNiO,DiTusa}. In this
material, the $S=1$ spins are formed by a strong ferromagnetic (FM)
Hund's coupling $J_H$ between the two active orbitals of
$\rm Ni^{2+}$. Theoretically, a number of interesting questions arises by
doping of a $S=1$ chain with mobile holes. The competition between FM
order induced by the double exchange mechanism \cite{DblExch} and AF
order can result in completely different magnetic properties, 
the spin gap can immediately be destroyed upon doping or might persist,
and finally which correlation function characterized by the single
correlation exponent $K_\rho$ dominates in the thermodynamic limit.

In a previous paper \cite{Spin1}, we have investigated the properties
of a doped spin-1 chain with a level difference between the two
electrons forming the $S=1$ spin. Such a situation is likely in a
system where the $S=1$ spins are formed by two electrons in different
orbitals of the same ion, as it is the case in $\rm Y_{2-x}Ca_xBaNiO_5$
\cite{Penc,Dagotto,Riera}.
Here we will concentrate on the effect of a level difference, by
comparing the above case with a system with electrons in two
equivalent levels. The latter situation is more likely in a
ladder-system. A few years ago, this problem has been studied by
Fujimoto and Kawakami in a weak coupling approach, where two
ferromagnetically coupled Hubbard-chains have been investigated and
compared to a Hubbard chain ferromagnetically coupled to a
Heisenberg-chain \cite{Fujimoto}. In their analysis it is found that
the model with two equivalent chains remains gapful, with a finite
string-order parameter and dominant CDW correlations ($K_\rho=1/2$). In
contrast, the model with a level-difference becomes gapless upon
doping and shows two gapless spin-modes. Similar conclusions have been
drawn by Nagaosa and Oshikawa in a semi-classical approach with two
coupled Hubbard chains \cite{Nagaosa}, and it is pointed out that for
low hole doping, the system remains gapful for equivalent doping on
the chains irrespective of FM or AF couplings on the rungs, whereas
the system becomes gapless upon doping for different hole
concentrations on both chains.  However, since both of these studies
are based on weak-coupling or mean-field approaches, it is not clear
whether the results are valid in this one-dimensional, strongly
correlated system and direct numerical evidence is needed. In this
Letter we would like to investigate and compare doping of a spin-$1$
chain with and without level difference in an unperturbative manner
free of approximations by the density matrix renormalization group
method (DMRG) \cite{White}.  This method is ideal for the study of
groundstate energies and equal-time correlation functions, and we
determine the correlation exponent $K_\rho$ and the string
order parameter.  Additionally we also calculate the magnetic
susceptibility at finite temperatures by the thermal DMRG (TDMRG)
method \cite{TDMRG,BiOTDMRG}.

The first model we consider is a model with equal particle concentrations
on both legs, and we call this the symmetric model in the following.
The $S=1$ spins are formed by two ferromagnetically coupled \tJ chains
with additional AF couplings between next-nearest neighbor sites 
on opposite legs. The individual terms which define the model are given
by hopping along the chain $i$
\[
H_{\text{kin}}^{(i)} = -t \sum_{j,\sigma} \cP \left(
        c^{\dagger}_{j,i,\sigma} c_{j+1,i,\sigma}
        +H.c.\right) \cP ,
\]
where the projection operator $\cP$ prohibits doubly occupied sites,
and $c_{j,i,\sigma}^\dagger$ is the particle creation operator on rung
$j$ and leg $i$ with spin $\sigma$. Further the AF couplings $J>0$ on
the chain $i$ between nearest neighbors and the diagonal couplings
$J_d>0$ are given by
\begin{eqnarray}
H_{\text{af}}^{(i)}& =& J \sum_{j} \left({\mathbf{S}}_{j,i} 
					{\mathbf{S}}_{j+1,i}
		-{1\over4}n_{j,i}n_{j+1,i}\right) \nonumber \\
H_{\text{diag}}& =& J_d \sum_{j,l} \left({\mathbf{S}}_{j,1} 
					{\mathbf{S}}_{l,2}
	-{1\over4}n_{j,1}n_{l,2}\right)
	\left( \delta_{l,j-1}+\delta_{l,j+1}\right) ,
\nonumber
\end{eqnarray}
where the notation is standard, the indices are the same as above, 
and $\delta_{j,l}$ denotes Kronecker's delta function. The strong Hund's 
coupling $J_H<0$ acts on the rungs
\[
H_{\text{FM}} = J_H \sum_{j} {\mathbf{S}}_{j,1} {\mathbf{S}}_{j,2}.
\]
Finally our Hamiltonian reads 
\[
H_{\text{sym}} = H_{\text{FM}} + H_{\text{diag}} + \sum_i \left(
	H_{\text{af}}^{(i)} + H_{\text{kin}}^{(i)} \right) .
\]
The second model contains a level difference between the two electrons
forming the $S=1$ spins and the lower band is localized. We will refer to this
model as the asymmetric model in the rest, and it consists of a 
\tJ chain ferromagnetically coupled to a $S=1/2$ HB chain, with additional
AF couplings between next-nearest neighbors on opposite legs. The Hamiltonian
is defined as
\[
H_{\text{sym}} = H_{\text{FM}} + H_{\text{diag}} +
		H_{\text{af}}^{(1)} + H_{\text{kin}}^{(1)}.
\]
For both models we restrict ourselves to parameter values with $J=J_d$
and $|J_H| \gg t,J,J_d$, and if not otherwise mentioned we set
$-J_H=10t=20J=20J_d$. In the DMRG calculations we use systems sizes of
up to $2 \times 256$ sites and up to 1300 states per system- and
environment-block. In the TDMRG we keep 80 states per block and use
finite Trotter-time steps of $\Delta \tau=0.2t$.

We have reported on the asymmetric model in a previous paper
\cite{Spin1}, and we briefly summarize the main results here. In that
system, each mobile hole creates a small FM cloud by the double
exchange mechanism (polaron). There is a very weak AF interaction
among the polarons giving the lowest lying, gapless spin-excitations.
However, the polarons are only a weak perturbation on top of the
underlying spin-liquid of the Haldane chain, and there is a second
energy scale in the spin-sector of the order of the Haldane gap of the
undoped system. This energy scale shows up in exponentially decaying
spin-correlations at short to intermediate distances and a finite
string-order. The dominant correlation in the thermodynamic limit is
given by $2k_F$ and $4k_F$ CDW order and the correlation exponent is
$K_\rho\approx 0.51$.

In the following we will perform a similar analysis for the symmetric
model and compare these findings with the asymmetric model.  We start
with the investigation of the spin gap. At half filling and $J_H\gg
J,J_d$, both models can be mapped to the Haldane $S=1$ chain with
effective couplings $J=J_{\text{eff}}^{\text{sym}}$ in the symmetric
case and $J_{\text{eff}}^{\text{as}}=3/4J$ in the asymmetric case.  By
finite size scaling we have numerically determined the spin gap from
$\Delta_s=\lim_{T\rightarrow 0} \Delta_{s}(L;N=Ln)$, where
$\Delta_{s}(L;N=Ln)=\Delta_{s}(L;N)=E_0(L;N;S^z=1)-E_0(L;N;S^z=0)$ and
$E_0(L;N;S^z)$ is the groundstate energy of the system with $N$
particles on $L$ sites and the total spin component along the
$z$-direction is $S^z$. The results are in excellent agreement with the
above mapping and we find $\Delta_{\text{spin}}^{\text{sym}}\approx
0.41(1) J_{\text{eff}}^{\text{sym}}=0.205(5)t$ for the symmetric model and
$\Delta_{\text{spin}}^{\text{as}}\approx 0.41(1)
J_{\text{eff}}^{\text{as}}\approx 0.154(5)t$ for the asymmetric model.
In agreement with refs.~\cite{Fujimoto,Nagaosa}, we find completely
different behavior for both models upon doping, as can be seen in
Table \ref{TabSpGap}. While the symmetric model retains a rather large
spin gap which decreases slightly upon doping ($n$ is the particle
density per single site), it is destroyed already for the smallest
hole density of $n_h^c=0.0625$ holes per site of the conduction band
for the asymmetric model.
\begin{table}[b]
\caption[*]{Finite size scaling of the spin gap $\Delta_{\text{spin}}$ for  
	$-J_H=10t=20J=20J_d$ and different values of the hole density
	$n_h=1-n$ for the symmetric model and hole density in the
	conduction band $n_h^c$ for the asymmetric model. Also listed
	is the spin-spin correlation length $\xi$.}
\label{TabSpGap}
\begin{tabular}{@{\hspace{\tabcolsep}\extracolsep{\fill}}lcccc}
\hline
$n_h$ & $\Delta^{\text{sym}}_{\text{spin}}$ & $\xi^{\text{sym}}$ &
	 $\Delta^{\text{as}}_{\text{spin}}$ & $\xi^{\text{as}}$ \\ \hline
0      &  $0.205(5) t$ &  $6.05(5)$ &  $0.1504(5) t$ & 	$6.05(5)$ \\
0.0625 &  $0.148(5) t$ & $6.1(2)$ &  0  & $7.9(2)$ \\
0.125  &  $0.120(5) t$ &  $6.7(2)$ & 0 & $11.2(2)$ \\ \hline
\end{tabular}
\end{table}

\myfig{Fig1}
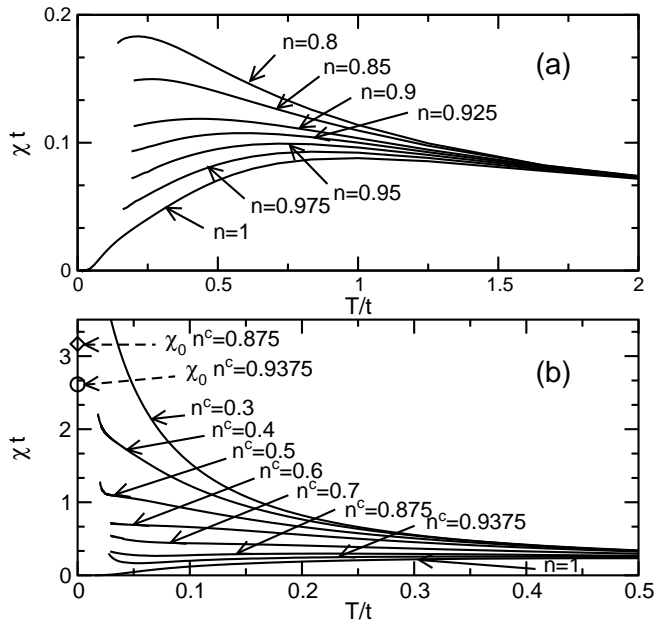
   \caption[*]{Temperature dependencies of the magnetic susceptibility $\chi$ 
	at different particle densities.
	Symmetric model (a) with $J=J_d=t/2$ and $J_H \rightarrow \infty$,
	and asymmetric model (b) with $-J_H=10t=20J=20J_d$.}
\label{FigSusc}
\end{figure}
At finite temperatures near the gap value $T\propto J/2$, the magnetic
susceptibility $\chi$ of both systems is strongly enhanced upon
doping, and we show the TDMRG results in Fig.~\ref{FigSusc}, where we
have set $J_F\rightarrow \infty$ for the symmetric model for numerical
reasons. This strong enhancement of $\chi$ is caused by the creation
of a small FM cloud around each mobile hole by the double exchange
mechanism, as the holes can gain kinetic energy by a FM alignment of
the neighboring spins. At still lower temperatures, the magnetic
susceptibility is suppressed for the symmetric model due to the spin
gap and we find $\chi \rightarrow 0$ for $T\rightarrow 0$. In
contrast, $\chi$ is finite and rather large at $T=0$ for the
asymmetric model, indicating the formation of larger FM
moments due to the double exchange mechanism (see ref. \citen{Spin1}).
However, the groundstate is spin-singlet for both models, as we have
tested with the DMRG method by the calculation of $\sigma(i,j)=S_i^z
S_j^z-1/2(S_i^+ S_j^-)$ which vanishes for a rotationally invariant
groundstate and the result is zero within the numerical precision of
the DMRG for this quantity for both models ($|\sigma(i,j)|<10^{-6}
\forall i,j$).

\begin{figure}[tb]
  \begin{center} 
      \includegraphics*[width=0.9\linewidth]{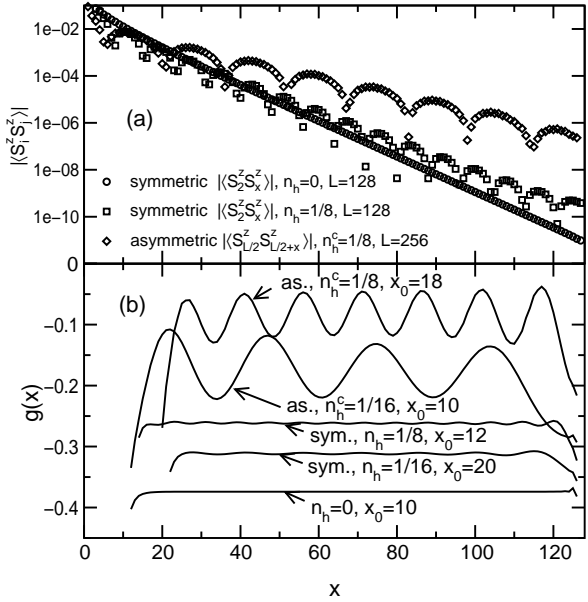}
  \end{center} 
   \caption[*]{Spin-spin correlations (a) for the symmetric and asymmetric
		 model with $-J_H=10t=20J=20J_d$, and 
		string-order $g(x)$ for both models with same parameters (b).}
\label{FigSpSpCorr}
\label{FigStringCorr}
\end{figure}
Complementary information about the magnetic properties is provided by
the spin-spin correlations shown in Fig.~\ref{FigSpSpCorr}(a).  
At half filling we find exponentially decaying correlations as
expected for gapful systems and the correlation length is $\xi\approx
6.03$ \cite{WhiteSpin1}. Also at finite doping, the spin-spin
correlations seem to decay exponentially for both systems. We have
determined $\xi$ by fitting to $S^z_iS^{z}_{i+x}\propto \cos(2k_F x)
e^{-x/\xi}$ and the results are listed in Table~\ref{TabSpGap}. In the
case of the symmetric model, the correlation length increases only
slightly upon doping to $\xi^{\text{sym}}\approx 6.7\pm0.2$ for
$n_h=1/8$, while the increase to $\xi^{\text{as}}\approx 11.2$ is
considerably larger for the asymmetric model at the same density. In
fact we expect a crossover to power-law decay at large distances in
the latter case (see ref.~\citen{Spin1}).
More insight about the spin-configuration can be gained from the
string-order parameter $g(x)=\langle (\sum_{i=1,2} S_{x_0,i}^z) \left(
\prod_{k=x_0+1, j=1,2}^{x-1} e^{i \pi S_{k,j}^z} \right) (\sum_{l=1,2}
S_{x,l}^z)\rangle$, \newline which reveals the hidden $Z_2 \times Z_2$
symmetry of the Haldane $S=1$ chain and quickly approaches the value
of $g(x)\approx -0.374$ for $x\gg 1$ in the undoped case
\cite{StringOrder}. From Fig.~\ref{FigStringCorr}(b) it can be seen that
$|g(x)|$ is reduced upon doping but remains finite for both models. A
finite string order parameter for the symmetric model has also been
obtained in ref.~\citen{Fujimoto}. Compared to the symmetric model, the
reduction of $|g(x)|$ upon doping is roughly three times larger for
the asymmetric model. In addition, the oscillations in $|g(x)|$ are a
strong amplification of the corresponding Friedel oscillations in the
charge density induced by the open boundary conditions. In fact, for
the asymmetric model, we again expect a crossover of $|g(x)|$ to
power-law decay at large distances after a sufficient number of
oscillations. By comparison, $|g(x)|$ is constant within $1.2\%$ for
the symmetric model apart from boundary effects, and we conclude that
its spin structure closely resembles that of the undoped case.

\begin{figure}[tb]
  \begin{center} 
      \includegraphics*[height=0.85\linewidth,angle=-90]{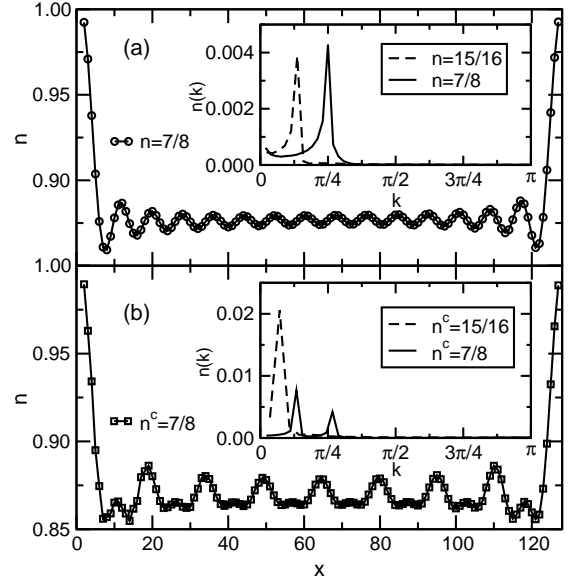}
	\caption[*]{Charge density for the symmetric model (a) at $n=7/8$ and 
	the symmetric model (b) at $n^c=7/8$, the parameters for 
	both models are $-J_H=10t=$ $20J=20J_d$ and the length is $L=128$.
	In the insets we show the corresponding Fourier transforms $n(k)$.}
\end{center}
\label{FigPairingCorr}
\end{figure}
Let us turn to the charge sector next and start with the above
mentioned Friedel oscillations in the charge density. In their Fourier
transforms we only find one peak at $k=2n\pi$ for the symmetric
model. This corresponds to the $2k_F$ CDW fluctuations, since the
lower band is fully occupied and hence there are $2(1-n)$ holes in the
upper band, giving $2k_F=(1+1-2(1-n))\pi\equiv 2n\pi$ for a large
Fermi volume. In the case of the asymmetric model we also find only
one peak at $k=n_h^c \pi$ for the smallest hole densities
$n_h^c=1/16$, and two peaks at $k=n_h^c \pi$ and $k=2n_h^c \pi$ for
the larger hole doping $n_h^c=1/8$. This is consistent with $2k_F$ and
$4k_F$ CDW fluctuations for a large Fermi volume including the
electrons of the lower orbital with $2 k_F=(n^c+1)\pi$ in that
case. Having identified these $2k_F$ and $4k_F$ fluctuations, we can
determine the correlation exponent $K_\rho$ by fitting to the Friedel
oscillations of an impurity potential. In the gapful case they are
given by $\delta_n(x)\propto C_1 \cos(2k_F x) x^{-K_\rho} + C_2
\cos(4k_F x) x^{-2K_\rho}$ and $\delta_n(x)\propto C_1 \cos(2k_F x)
x^{-(1+K_\rho)/2} + C_2 \cos(4k_F x) x^{-2K_\rho}$ in the gapless case
\cite{Fabrizio}. The correlation exponents thus obtained are
$K_\rho\approx 1.5\pm 0.05$ for the symmetric case at $n_h=7/8$
indicating dominant superconducting pairing correlations, and
$K_\rho\approx 0.51\pm 0.05$ for to asymmetric model at $n_h^c=1/8$,
giving dominant CDW correlations. For both cases, the sites near the
boundaries need to be discarded for the fit because of trapped states,
and the uncertainty stems from the fitting ambiguity. In the
symmetric model hole pairs are formed on the rungs in order to gain
the strong Hund's rule coupling $J_F$. The pair binding energy
$\Delta_{\text{pair}}\approx 2.29t$ in the low-doping region obtained
from $\Delta_{\text{pair}}=2 E_1-E_0-E_2$ is correspondingly large,
where $E_n$ is the groundstate energy with $n$ holes. This
pair-binding mechanism does of course not work for the asymmetric
model, but we still find a positive pair-binding energy
$\Delta_{\text{pair}}\approx 0.016t$ in that case. Pair formation in
that context has also been found in ref. \citen{Riera}. From a second
structure that develops in the hole-pockets at larger doping, we
expect that these pairs get spatially more extended upon doping.

The pairing correlations defined as
$P_{i}({\mathbf{f'}})P_{i+x}^\dagger({\mathbf{f}})$ with
$P_{i}^\dagger({\mathbf{f}})=\frac{1} {\sqrt{2}}(c_{i,\uparrow}^\dagger
c_{i+{\mathbf{f}},\downarrow}^\dagger \mp c_{i,\downarrow}^\dagger
c_{i+{\mathbf{f}},\uparrow}^\dagger)$ allow an independent calculation of
the correlation exponent $K_\rho$, and provide further information on
the form-factor $\mathbf{f}$ of the pairs. From the numerous
possibilities of the form-factor $\mathbf{f}$ for the symmetric model, we
restrict the discussion to those with the largest amplitudes in the
following.
\myfig{Fig4}
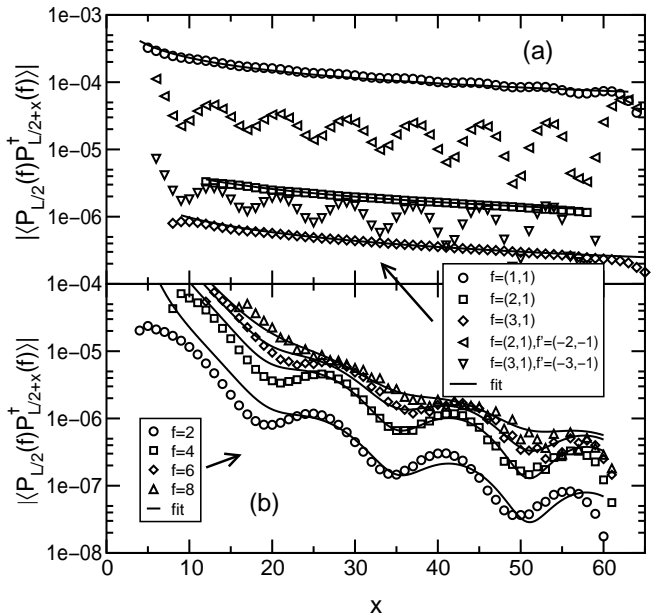
   \caption[*]{Singlet pairing correlations
   $P_{i}(f)P_{i+x}^{\dagger}(f')$ with $P_{i}^\dagger(f)=
   \frac{1}{\sqrt{2}}(c_{i,\uparrow}^\dagger c_{i+f,\downarrow}^\dagger
   -c_{i,\downarrow}^\dagger c_{i+f,\uparrow}^\dagger)$ 
	for (a) symmetric model at $n=7/8$ and (b) asymmetric model at
	$n^c=7/8$ and $-J_H=10t=$ $20J=20J_d$ for both models.}
\label{FigPairingCorr}
\end{figure}
At very short distances $d<5$ triplet pairing correlations for rung
pairs with ${\mathbf{f}}=(0,1)$ have the largest amplitudes, but as the
triplet correlation functions decay exponentially, the amplitudes for
singlet pairs formed on the rungs with an extension of the pair over
two to three rungs become larger at distances $d>5$. We show the
results for the singlet pairs with the largest amplitudes in
Fig.~\ref{FigPairingCorr}. These correlations decay as
$P_{i}(f)P_{i+x}^{\dagger}(f)\propto x^{-1/K_\rho}$ for a gapful
Luther-Emery liquid \cite{LutherEmery}, and surprisingly there are
almost no $2k_F$ fluctuations for the largest pairing correlations
despite the Friedel oscillations in the charge density. The
correlation exponent determined from the fit $K_\rho\approx 1.55 \pm
0.05$ is in excellent agreement with the result obtained from the
charge-density. For comparison, the correlation exponent obtained by
weak-coupling is $K_\rho=0.5$ \cite{Fujimoto}. The form factor of the
pairs with largest amplitudes ${\mathbf{f}}=(1,1), (2,1)$ and $(3,1)$ are
consistent with a $d_{x-y}$ symmetry analogue for a ladder with
vanishing ${\mathbf{f}}=(0,1)$ and $(1,0)$ amplitudes. For the
asymmetric model we obtain $K_\rho\approx 0.51 \pm 0.05$ for a gapless
Tomonaga-Luttinger liquid with $P_{i}(f)P_{i+x}^\dagger(f) \propto
A_0\ln(x)^{-1.5}x^{-1-1/K_\rho} +A_2\cos(2k_Fx)x^{-K_\rho-1/K_\rho}$
\cite{TomonagaLuttinger}
by simultaneously fitting to $f=2,4,6$, and $8$, again in excellent
agreement with the previous fit.

In conclusion we have studied the effect of a level difference on
doping of two-orbital chains. In addition to a finite spin gap for the
symmetric model and gapless excitations for the asymmetric model also
obtained by weak-coupling theories \cite{Fujimoto,Nagaosa}, we find
a new difference in the dominant correlation functions, which consist of
pairing correlations with $K_\rho \approx 1.5$ for the symmetric model
and CDW correlations with $K_\rho \approx 0.5$ for the
asymmetric model. Further the string-order parameter remains finite
for the symmetric model, however strongly reduced by the polarons in the
asymmetric model.

We wish to thank H. Asakawa and H. Tsunetsugu for valuable
discussions. The numerical calculations have been performed on
workstations at the ISSP. This work is supported by the ``Research for
the Future Program'' (JSPS-RFTF 97P01103) from the Japan Society
for the Promotion of Science (JSPS).


%
\end{document}